\documentstyle[preprint,aps,eqsecnum]{revtex}
\begin{document}
\draft
\title
{CALCULATION OF THE VACUUM ENERGY DENSITY  \\
 AND GLUON CONDENSATE WITHIN ZERO MODES \\
 ENHANCEMENT MODEL OF THE QCD VACUUM}
 
\author{V. Gogohia, Gy. Kluge and M. Priszny\'ak}
 
\address{ RMKI, Department of Theoretical Physics,
Central Research Institute for Physics, \\
H-1525,  Budapest 114,  P. O. B.  49,  Hungary}

\maketitle
 
\begin{abstract}
 The nonperturbative vacuum structure which emerges
from the zero modes enhancement (ZME) model of the true QCD
vacuum, appears to be well suited to describe
quark confinement, dynamical chiral symmetry breaking (DCSB),
current-effective (dynamical)-constituent, as
well as constituent-valence quark transformations, the
Okubo-Zweig-Iizuka (OZI) rule, dimensional transmutation, etc.
It is based on the solution to the Schwinger-Dyson
(SD) equation for the quark propagator in the infrared (IR)
domain.
The importance of the instanton-type fluctuations in the true QCD
vacuum for the ZME model is also discussed. This allows to
calculate new, more
realistic values for the vacuum energy density (apart from the
sign, by definition, the bag constant) and the gluon condensate.
Our numerical results for the gluon condensate for different
numbers of quark flavor $N_f$ are 2-3 times lager than it is
estimated in the QCD sum rules approach. This is in good agreement
with recent phenomenological estimates of this quantity.
\end{abstract}


\vfill
\eject

\section{Introduction}

    Today there are no doubts left that the dynamical mechanisms of
quark confinement and dynamical chiral symmetry breaking (DCSB)
are closely related to the complicated
topological structure of the QCD nonperturbative vacuum [1-3].
For this reason, any correct nonperturbative model of quark
confinement and DCSB necessary turns out to be a model of the true
QCD vacuum and the other way around.  Also it becomes clear that
the nonperturbative infrared (IR) divergenses, are closely
related, on one
hand, to the above mentioned nontrivial vacuum structure, on the
other hand, they are important as far as the large scale
behaviour of QCD is concerned [1-5]. If it is true that QCD
is an IR unstable theory (has no IR stable fixed point) then the
low-frequency modes of the Yang-Mills fields should
be enhanced due to the nonperturbative IR divergences. So the
gluon propagator can diverge faster than the free one at small
momentum, in accordance with  $D_{\mu\nu}(q) \sim (q^2)^{-2}$ at
small $q$, which describes the zero modes enhancement (ZME)
effect in QCD (see our preprint [6] and references therein).
If, indeed the low-frequency components of the virtual fields
in the true vacuum have a larger amplitude than those of the bare
(perturbative) vacuum [4], then the Green's function  for a
single quark should be reconstructed on the basis of this effect.
One of the main features of this reconstruction is, of course, the
correct treatment of this strong sigularity within the
distribution theory [7] that was precisely done in our previous
publications [6, 8]. Let us mentioned in this connection an
important observation that the correct treatment of this
behaviour within the distribution theory effectively transforms
the initial strong singularity ($q^{-4}$) into a Coulomb-like
behaviour ($ \sim q^{-2}$) in the intermediate and ultraviolet
(UV) regions, which is compatible with asymptotic freedom.
In fact, this transformation clearly shows that interaction
at short distances in QCD within the ZME effect is also completely
different from that of quantum electrodynamics (QED).
The possible ZME effect was our primary dynamical assumption.
We considered this effect as a very similar confining ansatz for
the full gluon propagator in order to use it as input
information for the quark , ghost Schwinger-Dyson (SD) equations as
well as for the corresponding Slavnov-Taylor (ST) identities
[6, 8, 9].
 
In our recent publication [10] and preprints [6, 11] the basic
chiral QCD
parameters (the pion decay constant in the current algebra (CA)
representation, $F_{CA}$, the quark
condensate, the dynamically generated quark mass, $m_d$, etc)
have been calculated from
first principles within the ZME model of quark confinement
and DCSB. As it was mentioned above it was based on
the solution to the SD
equation for the quark propagator in the infrared (IR) domain [6,
8-11]. At low energies QCD is governed by
$SU_L(N_f) \times SU_R(N_f)$ chiral symmetry
($N_f$ is the number of different flavors) and its dynamical
breakdown in the vacuum to the corresponding vectorial subgroup
[12]. Thus to understand chiral limit physics means to correctly
understand the dynamical structure of low energy QCD. A realistic
calculation of various physical quantities in this limit becomes
important. It is the goal of this work to generalize
our calculations to $N_f$ light quarks, as well as to discuss some
dynamical aspects of the ZME model which clearly shows the rich
possibilities of this effect in QCD.
 
There are only five independent quantities by means of which all
other chiral QCD parameters can be calculated in our model [10],
namely
 \begin{equation}
F^2_{CA} = {3\over {8 \pi^2}}k_0^2z_0^{-1}
             \int^{z_0}_0 dz \,{ zB^2(z_0,z) \over
             {\{zg^2(z) + B^2(z_0,z)\}}} ,
\end{equation}
\begin{equation}
m_d = k_0\bigl\{z_0 B^2(z_0,0)\bigr\}^{-1/2},
\end{equation}
\begin{equation}
{\langle \overline qq \rangle}_0 = -
{3\over {4\pi^2}}k_0^3z_0^{-3/2}{\int}^{z_0}_0 dz\,{zB(z_0,z)},
\end{equation}
\begin{equation}
\epsilon_q = - {3 \over {8 \pi^2}} k^4_0 z^{-2}_0
\int \limits_0^{z_0} dz\, z\, \{ \ln z\left[ z g^2(z) +
B^2(z_0, z)\right] - 2z g(z) + 2\},
\end{equation}
\begin{equation}
\epsilon_g = - {1 \over \pi^2} k^4_0 z^{-2}_0 \times
\left[ 18 \ln (1 + { z_0 \over 6})
       - {1 \over 2} z^2_0 \ln (1 + {6 \over z_0})
       - {3 \over 2} z_0 \right],
\end{equation}
where $\epsilon_q$ and $\epsilon_g$
describe a single confining quark and nonperturbative gluons (due
to the ZME effect) contributions to the vacuum energy density,
respectively. The explicit expressions for the solution of the
quark SD equation $g(z)$ and $B^2(z_0, z)$ are
\begin{equation}
g(z) =  z^{-2} [\exp{(-z)} -1 + z]
\end{equation}
and
\begin{equation}
B^2(z_0, z) =  3 \exp{(-2z)}
\int \limits_z^{z_0} {\exp{(2z')} g^2(z')\,dz'},
\end{equation}
respectively.
Thus our calculation scheme is self-consistent because we
calculate $n=5$ independent physical quantities by means of only
$m=2$ free parameters, which possess clear physical sense, so the
condition of self-consistansy $n>m$ is well satisfied. The mass
scale parameter $k_0$
characterizes the region where confinement, DCSB and
other nonperturbative effects are dominant while the
second indepent parameter $z_0$ is the constant of integration of
the quark SD equation. The
details of our scale-setting schemes for system (1.1-1.7) one may
find in Refs. [6, 10, 11] and numerical results in the case of a
single confining quark from Tables 1 and 2 therein.

\section{A possible dynamical picture of quark confinement and
        DCSB within the ZME effect in QCD }

 Let us make now a few detailed remarks (some of them are
necessary of semi-intuitive as well as of semi-speculative
character) sheding light
on our understanding of the actual dynamical mechanism of quark
confinement and DCSB might be interpreted with the help of the
possible ZME effect in QCD. Susskind and Kogut [3] have
noticed that "the absence of quarks and other colored objects
could only be understood in terms of an IR divergences in the
self-energy of a color bearing object". In our
approach gluons remain massless, only zero modes are enhanced.
We will discuss in more detail contributions
to the self-energy of the colored quark leading first to the
dynamical and then to constituent
quarks in the context of the ZME effect which is caused by the
nonperturbative IR divergences in the true vacuum of QCD.
 
In order to clarify the dynamical picture which lies at the
heart of our model, let us introduce, following
Mandelstam [4], two sorts of gluons. The actual (external) gluons
are those
which are emitted by a quark and absorbed by another one, while
the virtual (internal) gluons are those which are emitted and
absorbed by the same quark. Of course, both sorts of gluons are
not only free ones. All possible self-energy insertions are
aassumed to be taken into account as well.
At first sight  this separation seems to
be a simple convention but we will show below that it has a firm
dynamical ground, thus it makes our understanding of the above
mentioned picture more transparent.
 
 Let us consider now all the possible contributions to
the self-energy of a single quark. The most simplest one is
shown in Fig. 1. Let us recall that the same self-energy
diagram occurs also in the quantum electrodynamics (QED). In
contrast to QED, there is an infinite number of additional
contributions to the self-energy of a single quark because of
the non-abelian nature of QCD, i.e. because of the direct
interaction between virtual gluons which is absent in QED. Some
of these are shown
in Fig. 2. So, from the point of view of the contributions to the
self-energy of a single quark, the zero modes are indeed enhanced
in QCD in comparison with the electron self-energy in QED.
The self-interaction of virtual gluons alone removes
a single quark from the mass-shell, making it an effective
(dynamical in the chiral limit) object. This is the context of
the SD equation describing propagation of a single quark (at
large distances) in the true QCD vacuum.
 
 But this is not the whole story yet in QCD because up to now we
took
into account only contributions induced by the virtual gluons
alone.
The actual gluons emitted by one quark can contribute to the
self-energy of another quark and vice versa. The simplest diagrams
of this process are shown in Fig. 3. Moreover
contributions shown in Fig. 4
are also inevitable and they describe the process of
the convertation (transformation) of virtual gluons into
actual ones and the other way around.
Thus we consider diagrams, of these type, not as
corrections to the cubic and quartic gluon vertices but rather as
additional contributions to the self-energy of the quarks.
Contributions to the self-energy of each quark will be essentially
enhanced in the presence of another quark. In other words
each quark additionally enhances the interaction with the
vacuum (zero modes) of another quark.
Precisely this enhancement of the
zero modes by virtue of self-interaction of virtual (internal) and
actual (external) gluons is effectively correctly described
by the ZME model.
 
It is quite plausible that, at large
distances between quarks, actual gluons emitted by each quark do
repeatedly succeed to convert into virtual ones and vice
versa. This leads to a multiple enhancement of
the zero modes of each quark. Exactly these
additional contributions to the self-energy of each effective
quark makes it a constituent object in our model. The mass of
the constituent (heavy or light) quark becomes the sum of three
terms, namely
\begin{equation}
 m_q = m_{eff} + \Delta = m_0 + m_d + \Delta.
\end{equation}
All terms on the right
hand side have clear physical sense.
The first term is, obviously,  the current
mass of a single quark. The second one $m_d$ describes
contributions to the constituent quark mass induced by the
self-interaction of virtual gluons alone, while the third term,
$\Delta$, describes contributions to the constituent quark mass
which come from the process of the convertation of
actual gluons into virtual ones as it was discussed above.
This way our model provides a natural dynamical foundation of the
current-effective (dynamical)-constituent transformation [13] of
the quark degrees of freedom on the basis of non-abelian
character of the gluon fields. The existence of a nonzero
$\Delta$ is principal for our model but numerically it should not
be large, even for light quarks. Our intuition (based on the
obtained numerical results, see Tables 1 and 2 in Refs. [6, 10,
11]) tells us that it
is only of the order of a few per cent of the displayed
there values of the dynamically generated quark masses, $m_d$.
 
In reality the contributions are so mixed up that they
can not be separated from each other. There exists
an infinite number of
possible, topologically complicated, configurations of the vacuum
fluctuations of the non-abelian gauge (gluon) fields contributing
to the self-energy of each quark while making them constituent
objects. The true vacuum of QCD, however, is not settled by
these fluctuations alone, its structure is much
more richer [14] than that (see also the discussion below).
Certainly, a finite number of favorable, topologically distinct
vacuum configurations, which minimize the energy of the
bound states, should exist. It is hard to believe that in the real
word of four dimensions, the favorable topologically complicated
configurations are strings or planar ones. In this context, it is
important to comprehend that the linearly rising quark-antiquark
potential  at large distances, nicely showed by recent lattice
calculations [15], is not a privilege of the planar or string
configurations only. Though the above mentioned linear potential
does not contradict the ZME effect, nevertheless the potential
concept in general is a great simplification of the real dynamical
picture which emereges from our model. As it was underlined in our
papers [6, 8-11], the enhancement of zero modes necessary leads
to
full vertices while the potencial concept of the constituent quark
model (CQM) is based on point-like ones. Moreover, we think that
the potential concept has already played its useful role and now
should be retired from the scene like the Bohr orbits after the
creation of the true theory of atoms - quantum mechanics.
 
It is plausible that these energetically advantageous
configurations of vacuum fluctuations (leading to the
formation of the bound-states of the constituent quarks) occur at
a certain distance between constituent
quarks. They will be completely deformed (or even destroyed) if
one attempts to
separate the constituent quarks further from each other.
The nonperturbative vacuum of QCD is filled with
quark-antiquark virtual pairs which consist of various
components of quark degrees of freedom (light, heavy,
constituent, dynamical, etc).
This is an inevitable consequence of the ZME effect.
As a result of the above mentioned nontrivial topological
deformation, at least one quark loop will be certainly "cut".
We may, for convenience, think of this as such a topological
deformation which allows for quarks from
the loop to recombine with the initial constituent quarks. It is
evident that the breaking of the gluon line is not so important as
the above mentioned cut of the quark loop which is
equivalent to the creation of the corresponding
quark-antiquark pairs from the vacuum. The vacuum of QCD
will be immediately rearranged and, instead of
"free" constituent quarks, new hadron states will occur.
 
  At short distances the situation is completely different from
the above described. Indeed, in this case the actual gluons
emitted by each quark do not repeatedly succeed to convert into
virtual ones. So, at these distances, interaction between
quarks proceeds mainly through the exchange of actual gluons.
This means that the interaction of the constituent quarks with the
vacuum (i. e. contributions to its self-energy), due to the above
mentioned process
of convertation, is essentially decreased. So they become valence
quarks. The intensity of the process of convertation determines
the constituent-valence transformation. One may say that
the constituent (valence) quarks are "valence
(constituent)-in-being quarks". In other words, if the process of
convertation becomes stronger then the valence quark becomes
constituent and vice versa. If the process of convertation becomes
weaker then the constituent quark becomes valence, so hadron
becomes consisting of valence and sea quarks and mainly
actual gluons. Precisely this picture of hadrons emerges from deep
inelastic scattering experiments.

\subsection{The Okubo-Zweig-Iizuka selection rule}

  Any correct model of quark confinement should explain at least
qualitatively the famous Okubo-Zweig-Iizuka (OZI)
selection rule [16] since it reflects the unusual dynamics of
quarks inside hadrons, that, in turn, is closely related to the
QCD vacuum structure.
In the context of the ZME model it becomes
clear that the topological rearrangement of the vacuum by means
of the $direct$ annihilation of the initial (final)
constituent quarks, entering the same hadrons, is hardly
believable. In fact, what does the above mentioned $direct$
annihilation mean? This would mean
that the initial (final) constituent
quarks, emitting (absorbing) a number of
gluons, can annihilate with each other without the break-up of
the corresponding quark loops in our model. The initial (final)
constituent quarks always emit and adsorb
gluons in each preferable configuration.  Nothing interesting
should happen during these
processes. This is a normal phase of each preferable
configuration
and it describes only its trivial rearrangement. Any nontrivial
rearrangement of the vacuum can only begin with
cutting the quark loop. As it was mentioned above, this is
equivalent
to the creation of a quark-antiquark pair from the vacuum. Then
the annihilation of the initial (final) constituent quarks with
the corresponding quarks, liberated from the loop, becomes
possible. Diagramatically this looks like a $direct$ annihilation
(see Fig. 5).
The probability to create the necessary pair, in order to
annihilate the
initial (final) constituent quarks, is rather small, so, in
general, the annihilation channel must be suppressed.
It is much more
probable for the quarks liberated from the loop to recombine
with the initial (final) constituent quarks in order to generate
new hadron states.
In more complicated cases (when many quark loops are cut) the
annihilation of the initial (final) constituent quarks becomes
more
probable and this process will compete with the process of the
recombination of the initial (final) constituent and liberated-
from-the-vacuum quarks to generate new hadron states.
 
 The heavy constituent quarks inside hadrons
(for example, in $c \bar{c}$ systems) are much closer to each
other (the distance between them is of the order $m^{-1}_h \ll
m^{-1}_q$,
where $m_h$ and $m_q$ denote the masses of the heavy and light
constituent quarks, respectively) than their light counterparts.
However at short distances the interaction between them
proceeds mainly not through the vacuum
fluctuations but via the exhange of the actual gluons as this was
explained above. This means, in turn, that the number of virtual
loops which should be cut is small. Also the
probability to cut heavy quark loop, or equivalently to create a
heavy quark-antiquark pair from the vacuum, is much less
than to create, for example, a light pair.
The fluctuations
in the density of instantons (see next section) and condensates
during the vacuum's rearrangement also come into play.
Thus the process of the rearrangement
of the vacuum, leading to the transition between hadrons (their
strong decays) on the basis of the annihilation of the initial
(final) constituent quarks, as shown in Fig. 5, should be
suppressed in comparison with the process of the recombination
of the initial (final) constituent quarks with the quarks
liberated from the vacuum. This is shown in Fig. 6. Thus
in the case of heavy quarks  our qualitative dynamical
explanation of the OZI rule is in agreement
with its standard explanation which argues that the QCD coupling
constant becomes weak at short distances and suppresses
the annihilation channel. However this argument fails to explain
why the violation of the OZI rule for the pseudoscalar octet is
bigger than for the vector one.
 
   Let us analyse this problem in our approach.
Light constituent quarks inside pseudoscalar and vector mesons
are at relatively large distances ($\sim m^{-1}_q$ from each
other)
than their heavy counterparts, for example in $c \bar{c}$ systems.
This means that interaction between them is mediated mainly by the
vacuum fluctuations which provide plenty  of various
quark loops to be cut during the process of the vacuum
rearrangement. So the annihilation channel should not be
suppressed for these octets. Indeed, the violation of the OZI rule
in the pseudoscalar mesons is not small, but for the vector mesons
it is again small, i.e.  comparible to the violation in the
$c \bar{c}$ systems. Our model
provides the following explanation for this problem:
The same quark-antiquark pair in pseudoscalar and vector
mesons is in the same $S$-state. The only difference between
them is in the relative orientation of the quark spins.
Quark and antiquark spins are oriented
in the same direction in the vector mesons, while in the
pseudoscalar mesons their orientation is opposite. This
is schematically shown in Figs. 7 and 8.
For light constituent quarks spin effects become important, while
for heavy constituent ones such a relativistic effect as spin and,
in particular, its orientation is not so important. As it was
repeatedly
mentioned above, a nontrivial rearrangement of the vacuum in
our model always
starts from the cut of the quark loops. In order to analyse
the violation of the OZI rule, from the point of view of
annihilation of spin degrees of freedom, let us think
of quark loops as "spin loops". In pseudoscalar mesons at
least one spin-antispin liberated-from-the-vacuum-pair is needed
to annihilate the initial pair. This is schematically shown in
Fig. 7.
In vector mesons at least two spin-antispin
pairs are needed for this purpose, but, in addition, an
intermediate meson (exited) state certainly appears, see Fig. 8.
This means that the annihilation channel for the vector mesons,
unlike for the pseudoscalars ones, is suppressed.
It is worth noting that the OZI rule is a selection rule and it is
not a conservation law of some quantum number. So its breakdown is
always possible and suppressed processes may proceeds through the
appropriate intermediate states [17]. Exactly this is
shown in Fig. 8 schematically . All this explains the
violation of the OZI rule in the pseudoscalar channel in
comparison with the vector one in our model. For vector mesons the
decay $\phi \rightarrow 3 \pi, \rho \pi$, proceeding though the
annihilation channels, are suppressed in comparison with the decay
$\phi \rightarrow K^+ K^-$ which occurs via the recombination
chaneel while for pseudoscalar mesons, say, the decay
$\phi \rightarrow 3 \pi$ is not suppressed.
In order to confirm our qualitative explanation
of the OZI selection rule quantitatively, it is necessary to
calculate the strong decay widths of mesons in our model, which,
of course, is beyond the scope of the present paper.

\section{Instantons}

 The main ingredients of the QCD vacuum, in our model, are
quark and gluon condensates, quark-antiquark virtual pairs (sea
quarks) and self-interacting nonperturbative
gluons. The vacuum of QCD has, of course, much more remarkable
(richer) topological structure than this. It is a very
complicated medium and its topological complexity means that its
structure can be organized at various levels and it
can contain perhaps many other components [1, 14] besides
the above mentioned. There are a few models of the nonperturbative
vacuum of QCD which are suggestive of what a possible confinement
mechanism might be like (see recent paper [18] and references
therein). We will not discuss these models; let us only
mentioned that the QCD-monopole condensation model proposed
by t' Hooft and Mandelstam [19] within the
dual Ginzburg-Landau effective theory [20] also invokes the ZME
effect as well as the mechanism (classical) of the
confining medium recently suggested by Narnhofer and Thirring
[21]. Let us ask one of the main questions now.
 
 What is a mechanism like which initiates a topologically
nontrivial rearrangement of the vacuum?  It is
already known, within our model, that this may begin with the cut,
at least, of one
quark loop and therefore, at least, one quark-antiquark pair
emerges from the vacuum. Why can the quark loop be cut at all
and what prevents the quarks from the cut loop to
annihilate again with each other? The fluctuations in the
nonperturbative vacuum of QCD must exist which do these job.
This is an inevitable consequence of our model of the vacuum.
We see only one candidate for this role in four dimensional QCD,
namely instantons and anti-instantons and their interactions
[1, 14, 22].
 
Instantons are classical (Euclidean) solutions to the
dynamical equation of motion of the nonabelian gluon fields
and represent topologically nontrivial fluctuations of these
fields. Self-interaction of gluons should be important
for the existence of the instanton-like fluctuations
even at classical level.
In the random instanton liquid model (RILM) [23], light quarks
can propagate over large distances in the QCD vacuum by simply
jumping from one instanton to the next one. In contrast, in
our model the propagation of all quarks is determined by the
corresponding SD equations (due to the ZME effect) so that they
always remain off mass-shell. Thus we do not need the
picture of jumping quarks. As opposed to the
RILM, we think that the main role of the instanton-like
fluctuations is precisely to prevent the quarks and gluons from
freely propagating in the QCD vacuum. Running against
instanton-like fluctuations, the quarks undergo difficulties in
their propagation in the QCD vacuum which, as was mentioned above,
is a very complicated inhomogenious medium.
At some critical value of the instanton density the
free propagation of the virtual quarks from the loops
become impossible so they never annihilate again with
each other. Obviously, this is equivalent to the creation of the
quark-antiquark pairs from the vacuum. From this
moment the nontrivial rearrangement of the vacuum may start.
The role
of the instanton-like fluctuations appears to be "cutting"
the quarks loops and preventing them from the immediate
annihilation of quarks and antiquarks liberated from the loops.
Thus liberated from loops quarks may, in principle, recombine or
even annihilate with initial constituent quarks to produce new
hadron states. In this way precisely instantons may promote
transitions between hadrons, i.e. they destabilize energetically
advantageous (dominant) configurations of the vacuum fluctuations
which lead to hadron states. One of the main features of the
instanton-induced effects is tunneling between topologically
distinct vacuums in Minkovski space [1].
Our understanding of their role in the QCD vacuum structure is in
agreement with this.
 
Being classical (not quantum!) fluctuations, instantons
can cut the quark loops in any points even in the quark-
gluon vertices. A simple cut of the quark loops, as it was
repeatedly emphasized above, is equivalent to the creation of the
corresponding quark-antiquark pairs from the vacuum. Let
us ask now what happens if all the external (actual) gluon lines
will be cut from the quark loops by the instantons. Well, in this
case each quark loop becomes a closed system. Because of the
vacuum pressure they immediately collapse and one obtains nothing
else but quark condensate if, of course, all internal gluon lines
can
be included into the quark self-energy. Another scenario is also
possible when not all the internal gluon lines can be included
into the quark or gluon self-energy. In general, the presence of the
internal gluon lines in the vacuum diagrams prevent them from
collapsing (because they counterbalance the vacuum pressure) and
consequently they should contribute to the vacuum energy density.
 
It would be suggestive to conclude that the same mechanism
works in order to produce gluon condensates despite the much more
complicated character of the gluon self-interactions. But this
is not the case indeed, since there is a principal difference
between quark and gluon condensates. The former ones do not
contribute to the vacuum energy density (in the chiral limit,
see below) and, in this sense, they
play the role of some external field. While the latter ones, as
was shown first in Ref. [24], are
closely related to the vacuum energy density. Like we noted
previously, nothing interesting should be happen if the instantons
cut
the gluon line in one point. Let us imagine that many gluon lines
will be cut by the instantons in many points. The vacuum
becomes filled up with gluon pieces (segments) which, due to
the existence
of the gluon strong self coupling, can recombine, in principle,
as some colour singlet bound-states --
gluonia or glueballs. Unlike quark condensates,
glueballs should have internal pressure because of the strong
self-interaction of composite gluon segments which prevents them
from collapsing. In turn, this means that the glueballs should be
heavy enough.
 
  The pseudoscalar mesons (consisting of light quarks)
are Nambu-Goldstone (NG) states so their
masses remain zero in the chiral limit even in the presence of
DCSB. From our model it follows that the existence of the
instanton-type fluctuations in the true vacuum promote strong
meson decays by preventing quarks and gluons from freely
propagating in it. So one can conclude in that
instanton-type fluctuations should be totally suppressed in this
case in
order to provide stability for the massless NG states since
massless particles cannot decay. This feature of the
instanton physics in the massless quarks case was discovered
by t' Hooft [25]. In the presence of DCSB, however, the
instanton-type fluctuation are restored [14, 26], but
the contribution of the instanton component to the vacuum energy
density, $\epsilon_I$, still remains small. So the dilute gas
approximation for the
instanton component seems to be relevant in this case. Not going
into details of the instanton physics (well described by Callan,
Dashen and Gross in Ref. [14]), let us only emphasize that at
short
distances the density of small size instantons should  rapidly
decrease and conversely increase at large distances where large
scale instantons, anti-instantons and their interactions also come
into play. Otherwise it would be
difficult to understand the role which we would like to assign
to the instantons in our model. This is in agreement
with the behaviour of the instanton component of the QCD
vacuum at short and large distances described in the above
mentioned paper [14], as well as
with our understanding of the actual dynamical mechanism of quark
confinement and DCSB. It is possible to say that we treat the
instanton component of
the QCD vacuum not as a "liquid" but rather as a "forest" in which
instantons and anti-instantons are considered as "trees" with
$SU_c(3)$ orientation, position and scale size.

\section{Calculation of the vacuum energy density }

The nontrivial rearrangement of the vacuum
can start only when the density of instantons achieves
some critical values, different for all distinct vacuums.
For this reason, despite being a classical phenomena,
instantons should nevertheless contribute to the vacuum
energy density through the above mentioned quantum tunneling
effect which is known to lower the energy of the ground-state.
As it was noticed above, $\epsilon_I$
should be small for light quarks with dynamically generated masses.
However, the same conclusion seems to be valid for heavy quarks as
well in our model. Heavy quarks are at short distances from each
other (in mesons) at which the nonperturbative effects, such as
instantons and enhancement of zero modes, are suppressed. Thus the
dilute gas approximation for the instanton component seems to be
applicable to light quarks with dynamically generated masses as
well as to heavy quarks. In this case it is worth
assuming, following the authors of Ref. [24], that light and
heavy quarks mutch smoothly (in our model this is almost
inevitable consequence). In the above mentioned RILM [23] of the
QCD vacuum, for a dilute ensemble, one has
\begin{eqnarray}
\epsilon_I &=& - (1/4) (11 - {2 \over 3} N_f) \times 1.0 \ fm^{-4}
\nonumber\\
&=& - (0.00417 - 0.00025 N_f) \ GeV^4.
\end{eqnarray}
For $N_f=3$ it coinsides with the estimate of the QCD sum rules
approach on account of the phenomenological value of the gluon
condensate [24] (via the trace anomaly relation, see below).
It decreases with increasing $N_f$ that defies a physical
interpretation of the vacuum energy density as the energy per unit
volume. That is why the value of the vacuum energy density at the
expense of the instanton contributions alone is at least not
complete.
 
In QCD the vacuum energy density (at the fundamental quarks
and gluons level) should be calculated through the effective
potential method for composite operators [27] (for review see
Refs. [28]), since in the
absence of the external sources the effective potential is nothing
but the vacuum energy density. This method gives it in terms
of loop expansion in powers of Plank constant. Using it
we have already calculated [6, 10] the
contributions to the vacuum energy density of confining quarks
with dynamically generated masses, $\epsilon_q$ (1.4), and
nonperturbative gluons, $\epsilon_g$ (1.5), due to enhancement of
zero modes (both at log-loop level). Thus the value of the vacuum
energy density, as given by the ZME model, is
\begin{equation}
\epsilon_{ZME} = \epsilon_g + N_f \epsilon_q,
\end{equation}
where we introduced the dependence on the number of
different quark flavors $N_f$ since $\epsilon_q$ itself is the
contribution of a single confining quark. Numerically it is
\begin{equation}
\epsilon_{ZME}= - (0.0016 + 0.0015 N_f) \ GeV^4.
\end{equation}
(For numerical values of each component $\epsilon_g$ and
$\epsilon_q$ in calculation scheme A see our paper [10] (Table 1)).
 
 However, neither contribution (4.1) nor (4.3) is complete. It was
already explained in detail above why the instanton-type
fluctuations are needed for the ZME model. In order to get a more
realistic value of the vacuum energy density, let us add
$\epsilon_I$, as given by Eq. (4.1), to the ZME model value (4.3),
i. e. let us put that the total (t) vacuum energy density at
least is
\begin{equation}
\epsilon_t = \epsilon_I + \epsilon_{ZME} = \epsilon_I + \epsilon_g
+ N_f \epsilon_q,
\end{equation}
and numerically it is
\begin{equation}
\epsilon_t = - (0.00577 + 0.00125 N_f) \ GeV^4.
\end{equation}
Note, the dependence on $N_f$ now becomes correct.
The above described components produce the main ((leading)
contribution t the vacuum energy density. The next-to-leading
contributions (as given by the effective potential for composite
operators at two-loop level [27]) are $h^2$-order, where $h$ is
the above mentioned Plank constant. Thus they are suppressed at
least by one order of magnitude in comparison with our values (4.5).
 
In the calculation scheme B bounds for the total vacuum energy
are
\begin{equation}
- 0.00638 - 0.00111 N_f \leq \epsilon_t \leq - 0.00461 -
0.00145 N_f
\end{equation}
in units of $GeV^4$.
(For numerical values of each component $\epsilon_g$ and
$\epsilon_q$ in calculation scheme B see our paper [10] (Table 2)).

\section{Calculation of the gluon condensate}

The vacuum energy density
is important in its own right as the main characteristics of the
nonperturbative vacuum of QCD. Futhermore it assists in
estimating such an important phenomenogical parameter as
the gluon condensate, introduced within the QCD sum rules approach
to resonance physics [24]. Indeed, because of the Lorentz
invariance,
\begin{equation}
\langle{0} | \Theta_{\mu\mu} | {0}\rangle = 4 \epsilon_t
\end{equation}
holds where $\Theta_{\mu\mu}$ is the trace of the energy-momentum
tensor and $\epsilon_t$ is the sum of all possible contributions
to the vacuum energy density. According to QCD the famous
trace anomaly relation [29] in the general case (nonzero "bare"
quark masses $m_f$) is
\begin{equation}
\Theta_{\mu\mu} = {\beta(\alpha_s) \over 4 \alpha_s}
G^a_{\mu\nu} G^a_{\mu\nu} + \sum_f m_f \overline q_f q_f.
\end{equation}
($G^a_{\mu\nu}$ being the gluon field strength tensor).
The function $\beta(\alpha_s)$, up
to terms of order $\alpha^3_s$, is [1]
\begin{equation}
\beta(\alpha_s) = -(11 - {2 \over 3} N_f) { \alpha^2_s \over 2 \pi}
 + O(\alpha^3_s).
\end{equation}
Sandwiching Eq. (5.2) between vacuum states and on account of
relations (5.1) and (5.3), one obtains
\begin{equation}
\epsilon_t = - (11 - {2 \over 3} N_f)
 {1 \over 32}
\langle{0} | {\alpha_s \over \pi} G^a_{\mu\nu} G^a_{\mu\nu} | {0}\rangle
+ {1 \over 4} \sum_f m_f
\langle{0} | \overline q_f q_f | {0}\rangle,
\end{equation}
where
$\langle{0} | \overline q_f q_f | {0}\rangle$ is the quark
condensate and
$\langle{0} | {\alpha_s \over \pi} G^a_{\mu\nu} G^a_{\mu\nu} |
{0}\rangle$ is nothing but the gluon condensate [24].
The weakness of this derivation is, of course, relation (5.3)
which holds only in the perturbation theory. In any case, it would be
enlightening to numerically estimate the gluon condensate with the
help of relation (5.4).
 
From Eqs. (4.5) and (5.4) in the chiral limit ($m_f=0$) it finally
follows
\begin{equation}
\langle{0} | {\alpha_s \over \pi} G^a_{\mu\nu} G^a_{\mu\nu} |
{0}\rangle = {0.00577 + 0.00125 N_f \over 0.344 - 0.021
 N_f}  \ GeV^4.
\end{equation}
Numerically our values are:
\begin{eqnarray}
\langle{0} | {\alpha_s \over \pi} G^a_{\mu\nu} G^a_{\mu\nu} |
{0}\rangle (N_f=0) &\simeq& 0.01677 \ GeV^4, \nonumber\\
\langle{0} | {\alpha_s \over \pi} G^a_{\mu\nu} G^a_{\mu\nu} |
{0}\rangle (N_f=1) &\simeq& 0.0217 \ GeV^4, \nonumber\\
\langle{0} | {\alpha_s \over \pi} G^a_{\mu\nu} G^a_{\mu\nu} |
{0}\rangle (N_f=2) &\simeq& 0.0274 \ GeV^4, \nonumber\\
\langle{0} | {\alpha_s \over \pi} G^a_{\mu\nu} G^a_{\mu\nu} |
{0}\rangle (N_f=3) &\simeq& 0.034 \ GeV^4.
\end{eqnarray}
Thus our values of the gluon condensate (5.5-5.6)
are of factor of 2-3 larger than the value phenomenologically
estimated from the QCD sum rules approach [24], namely
\begin{equation}
\langle{0} | {\alpha_s \over \pi} G^a_{\mu\nu} G^a_{\mu\nu} |
{0}\rangle \simeq 0.012 \ GeV^4.
\end{equation}
There exist already phenomenological estimates of the gluon
condensate pointing out that its so-called standard value,
which is determined by the instanton-type fluctuations alone
(5.7), is too small. It was
pointed out (perhaps first) in Ref. [30] (see also Ref. [31])
that QCD sum rules
substantially underestimated the value of the gluon condensate
about of factor of 2-3. Our numerical results (5.5-5.6) are
in good agreement with these estimate. The most recent
phenomelogical calculation of the gluon condensate is given by
Narison in Ref. [32], where a brief review of many previous
calculations is also given. His analysis leads to the update
average value as
\begin{equation}
\langle{0} | {\alpha_s \over \pi} G^a_{\mu\nu} G^a_{\mu\nu} |
{0}\rangle = (0.0226 \pm 0.0029) \ GeV^4.
\end{equation}
Thus our results (5.5-5.6), calculated from first principles on
the basis of the ZME model, are in good agreement with
phenomenologically estimated values of the gluon condensate
summarized in the Narison's paper (see Table 2 in Ref. [32]).
 
In the calculation scheme B bounds for the gluon condensate on
account of Eqs. (4.6) and (5.4) become
\begin{equation}
{0.00461 + 0.00145 N_f \over 0.344 - 0.021 N_f} \leq
\langle{0} | {\alpha_s \over \pi} G^a_{\mu\nu} G^a_{\mu\nu} |
{0}\rangle \leq {0.00638 + 0.00111 N_f \over 0.344 - 0.021
 N_f} GeV^4
\end{equation}
in units of $GeV^4$.
Numerically these bounds are:
\begin{eqnarray}
0.0134 \leq \langle{0} | {\alpha_s \over \pi} G^a_{\mu\nu}
G^a_{\mu\nu}
| {0}\rangle (N_f=0) \leq 0.0185,  \nonumber\\
0.0187 \leq \langle{0} | {\alpha_s \over \pi} G^a_{\mu\nu}
G^a_{\mu\nu}
| {0}\rangle (N_f=1) \leq 0.0232,  \nonumber\\
0.0248 \leq \langle{0} | {\alpha_s \over \pi} G^a_{\mu\nu}
G^a_{\mu\nu}
| {0}\rangle (N_f=2) \leq 0.0285,  \nonumber\\
0.032 \leq \langle{0} | {\alpha_s \over \pi} G^a_{\mu\nu}
G^a_{\mu\nu}
| {0}\rangle (N_f=3) \leq 0.0345.
\end{eqnarray}
These bounds are also in agreement with values of the gluon
condensate summarized in the above mentioned Narison's paper
(Table 2 in Ref. [32]).
The lower and upper bounds for the gluon condensate
\begin{equation}
0.04 \leq \langle{0} | {\alpha_s \over \pi} G^a_{\mu\nu}
G^a_{\mu\nu} | {0}\rangle \leq 0.105 \ GeV^4,
\end{equation}
recently derived from the families of $J/ \Psi$ and $\Upsilon$
mesons in Ref. [33] slightly and substantially overestimate
our lower and upper bounds (5.10), respectively.
The tendency, however, to increase the value
of the gluon condensate is correct.

\section{The bag constant}

  A nontrivial relation between our
model, on one hand, and the bag [34] and string [35]
models, on the other hand, would not be surprised.
In this connection some dynamical aspects of our model
should be underlined. From the above consideration it follows
that,
from a dynamical point of view, maybe the ZME effect does
not lead to string configuratons of flux tube type between quarks.
Nevertheless, there is no doubt that this dynamical process works
like a
string preventing quarks to escape from each other. It takes place
in the finite volume of the QCD vacuum but it does not
require the introduction of an explicit surface. The finiteness
of the cut-off $z_0$ results in unphysical singularities (at this
point $z_0$) of
the solutions to the quark SD equation which are due to inevitable
ghost degrees of freedom in QCD. This has nothing to do with the
bag fixed boundary. Numerically it depends on a scale at which
nonperturbative effects become essential in our model. We treat
a hadron as a dynamical process which takes place in some
finite volume of the vacuum rather than as an extended object with
an explicitly fixed surface in the vacuum. The ZME model
remains a local field theory. However, the existence of
the vacuum energy per unit volume -- the bag constant $B$ -- is
important in our model as well. The inward positive pressure $B$
counterbalances the vacuum energy density needed for
generating the vacuum fluctuations, inspired  by the
enhancement of the zero modes in our model, i. e. the sum of the
bag constant and the nonperturbative vacuum energy density must
be zero. We
consider the bag constant as a universal one which
characterises the complex nonperturbative structure of the QCD
vacuum itself and it does not depend on the hadron matter.
 
The bag constant is defined as the difference between the energy
density of the perturbative and the nonperturbative QCD vacuums.
We normalized the perturbative vacuum to zero [6, 10], so in
our notations the bag constant becomes
\begin{equation}
B = - \epsilon_t = (0.00577 + 0.00125 N_f) \ GeV^4,
\end{equation}
on account of Eq. (4.5). So our predictions for a more realistic
values of the bag constant are:
\begin{eqnarray}
B (N_f=0) \simeq 0.006 \ GeV^4 \simeq (278 \ MeV)^4 \simeq 0.78 \
GeV/fm^3, \nonumber\\
B (N_f=1) \simeq 0.007 \ GeV^4 \simeq (290 \ MeV)^4 \simeq 0.91 \
GeV/fm^3, \nonumber\\
B (N_f=2) \simeq 0.008 \ GeV^4 \simeq (300 \ MeV)^4 \simeq 1.04 \
GeV/fm^3, \nonumber\\
B (N_f=3) \simeq 0.0095 \ GeV^4 \simeq (312 \ MeV)^4 \simeq 1.24 \
GeV/fm^3.
\end{eqnarray}
It has been noticed in [36] that
noybody knows yet how big the bag constant might be, but generally
it is thought it is about $1 \ GeV/fm^3$. The predicted value
for the most relevant physical case of $N_f=2$ is in fair
agreement with this expectation.
 
In the calculation scheme B bounds for the bag constant on account
of Eq. (4.6) are
\begin{equation}
 0.00461 + 0.00145 N_f  \leq B \leq 0.00638 + 0.00111 N_f
\end{equation}
in units of $GeV^4$.

\section{discussion}

 Let us make now a few things perfectly clear. It makes sense to
underline once more that the vacuum energy density $is \ not$
determined by the trace anomaly relation (5.4). It is much more
fundamental quantity than the gluon condensate and the gluon
condensate itself is determined by the vacuum energy density
via the trace anomaly relation (5.4) in the chiral limit. As was
explained above, the vacuum energy density in QCD is to be
calculated completely independently from the gluon condensate.
The gluon or quark condensates may or may not exist but the vacuum
energy density as energy per unit volume always exist. In QCD at
quantum level it should be calculated (as it was pointed out above)
by the effective potential method for composite operators [27].
It gives the vacuum energy density in the form of loop expansion
where the number of the two-particle irreducible vacuum loops
(consisting mainly of confining quark and nonperturbative gluon
degrees of freedom) is equal to the power of the Plank constant.
So the vacuum energy density at quantum level to leading
order becomes in general $\epsilon_g + N_f \epsilon_q
+ O(h^2)$. We beleive that the ZME model correctly reproduces
these
contributions to the vacuum energy density (see Eqs. (4.2-4.3)).
 
But this is not the whole story in QCD. The instanton-type
fluctuations also exist in the nonperturbative vacuum of QCD and
they contribute to the vacuum energy density as well. This
contribution, however, is a contribution at classical level.
So the vacuum energy density becomes the sum of all possible
contributions and only this sum determines the gluon condensate in
the chiral limit via the trace anomaly relation. That is why the
realistic value of the gluon condensate is approximately 2-3
times larger than it is estimated in QCD sum rules.
 
  As it was already mentioned above, a nontrivial rearangement of
the QCD true vacuum will occur if the density of the
instanton-like fluctuations achieves some critical value. This
critical value can be reached when $ - \epsilon_I \gtrsim
- ( \epsilon_g + \epsilon_q)$, i. e. when at least one sort
of quark flavors is
presented in the QCD vacuum. Using our numerical results (4.3) as
well as (4.1) for $N_f=1$,
it is easy to see that this condition is nicely satisfied, namely
$ 0.00395 \ GeV^4 \gtrsim 0.0031 \ GeV^4$. Within the ZME
model, this clearly shows that Shuryak [23] apparently correctly
suggested the average
distance between instantons as $1 \ fm$ and consequently the
density of instantons as $1 \ fm^{-4}$ although he started from
the underestimated value of the gluon condensate which was
phenomenologically given by the QCD sum rules approach (5.7).
Then he used the
trace anomaly relation (5.4) in the chiral limit for $N_f=3$ and
thus reproduced
the value of the vacuum energy density (4.1) which is due to
the instanton contributions alone. At that moment noybody knew
haw to calculate the vacuum energy density at quantum level and
the trace anomaly relation was a uniqe way to estimate it.
 
  Qiute recently in quenched ($N_f=0$) lattice QCD by using
the so-called "cooling" method the role of the instanton-type
fluctuations in the QCD vacuum was investigated [37]. In
particular, it was concluded that the instanton density should be
$n = (1 + \delta) \times fm^{-4}$, where $\delta$ was estimated as
$\delta \simeq 0.3 - 0.6$ depending on cooling steps.
So this enhancement in the density of instantons leads to the
enhancement of the vacuum energy density due to the instantons
contributions alone, leaving contributions from other components
unchanged, of course. Here two scenario are possible. First, this
really takes place and Eq. (4.1) should be improved on account of
the above mentioned enhancement in the density of instantons.
Let us argue, however, that Eq. (4.1) still remains valid and the
enhancement of the vacuum energy density is due to the
nonperturbative gluons contributions which apparently also can not
be removed by cooling method from the QCD vacuum. Indeed, the
additional contribution numerically is
\begin{eqnarray}
\epsilon_{\delta} &=& - (11/4) (0.3-0.6) \times \ fm^{-4}
 \nonumber\\
&=& - (0.00125 - 0.0025) \ GeV^4.
\end{eqnarray}
The average between these values $\epsilon_{\delta} \simeq 0.0019
\ GeV^4$ is rather close to our value for $\epsilon_g \simeq
0.0016 \ GeV^4$ (see Eq. (4.3)). If this is so then there is no
need to change key parameters of RILM [23] and perhaps of
the interacting instanton liquid model (IILM) [38].
What is necessary indeed is to take into account other possible
contributions to the vacuum energy density.
 
Thus one may conclude in that Eq. (4.1) rather correctly gives the
contribution of the instanton component to the
total vacuum energy density while the value of the gluon
condensate is precisely determined  by the total vacuum energy
density. This is the reason (as was mentioned above) why the
realistic value of the gluon
condensate 2-3 times larger than the instanton component alone can
provide. We have shown that additional contribitions to the vacuum
energy density and consequently to the gluon condensate are due to
the nonperturbative gluons and confining quarks on the basis of
the ZME model of the QCD true vacuum. Our numbers for the
gluon condensate's values
(obtained from first principles) are in good agreement
with recent phenomenological estimates summarized by Narison
in Ref. [32].
 
  One of the authors (V.G.) is grateful to N.B.Krasnikov for
useful discussions and remarks.

\vfill
\eject

 \vfill
 \eject

\begin{figure}

\caption{The simplest contribution to the quark
self-energy induced by a virtual (internal) gluon. }
 
\bigskip
 
\caption{The simplest contributions to the quark
self-energy induced by the self-interactions of virtual (internal)
  gluons. }
 
\bigskip
 
\caption{The simplest contributions to the quark
self-energy induced by an actual (external) gluons emitted by
another quark.}

\bigskip
 
\caption{The simplest contributions to the quark
self-energy due to the processes of the convertation
(transformation) of the virtual gluons into the actual ones and
vice versa.}
 
\bigskip
 
\caption{ The quark diagram for the decay of the $J/ \Psi$ meson.
The dashed line schematically shows the $c \bar{c}$ pair emerged
from the nonperturbative QCD vacuum. The true number of
intermediate
gluons is dictated by colour and charge conservation.}
 
\bigskip
 
\caption{ The quark diagram for the decay of the $\Psi''$ meson.
The dashed line schematically shows the $u \bar{u}$ pair emerged
from the nonperturbative QCD vacuum. }
 
\bigskip
 
\caption{ The quark diagram for the decay of the pseudoscalar (P)
particle. The arrows show the reciprocal orientation of the
spins. The dashed line schematically shows the pair emerged
from the QCD vacuum. The true number of intermediate
gluons is dictated by colour and charge conservation.}
 
\bigskip
 
\caption{ The quark diagram for the decay of the vector (V) meson.
 The arrows show the reciprocal orientation of the spins.
The dashed lines schematically show the pairs emerged
from the QCD vacuum. The true number of intermediate
gluons is dictated by colour and charge conservation.
An intermediate meson (exited) state is inevitable.}
\end{figure}

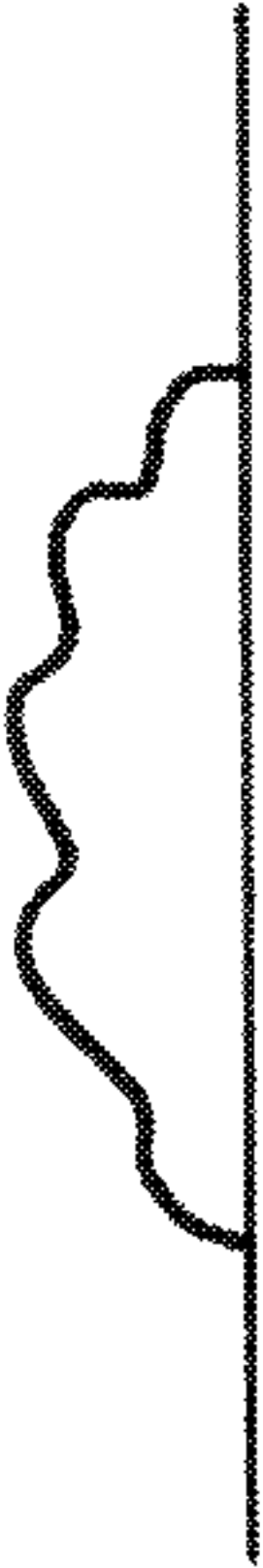
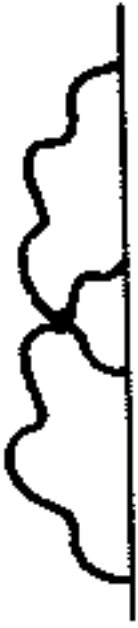
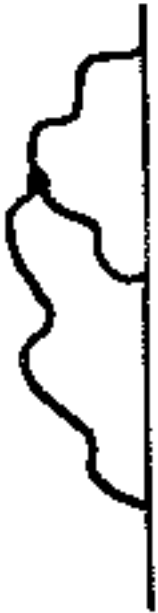
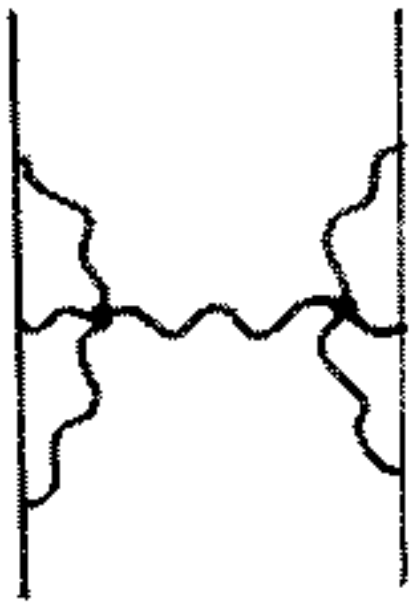
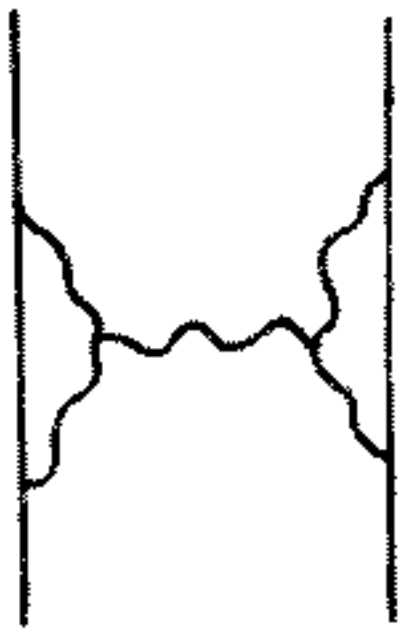
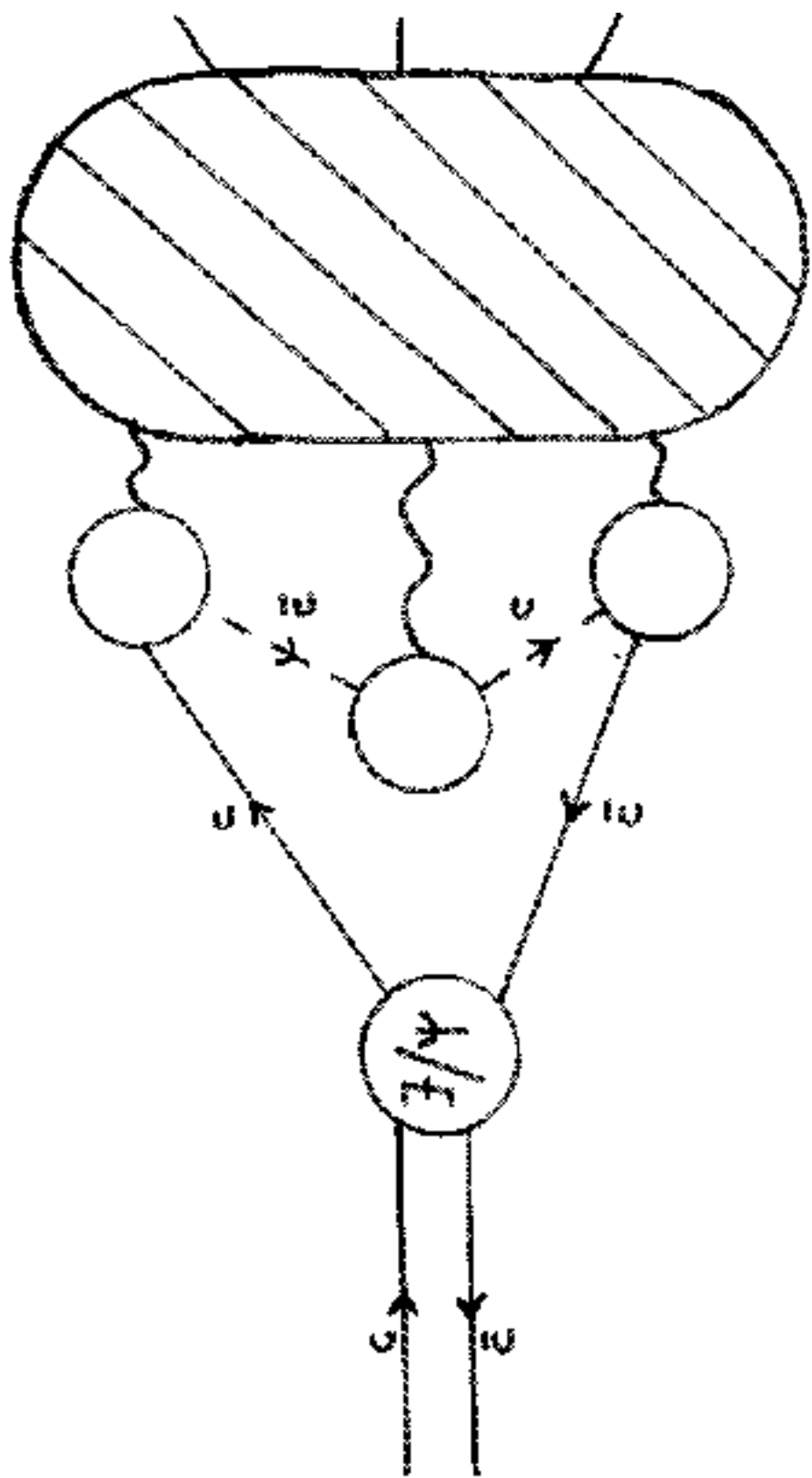
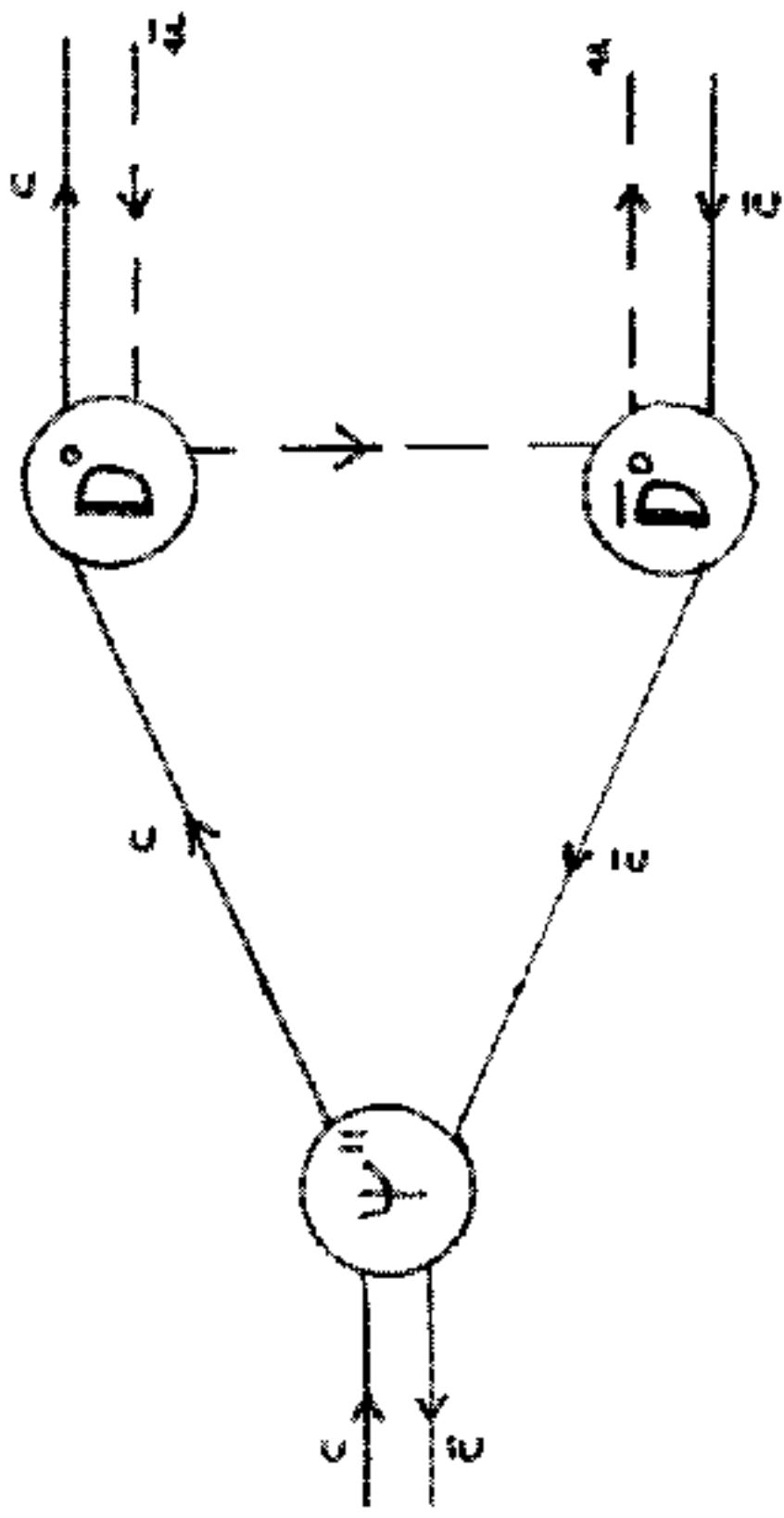
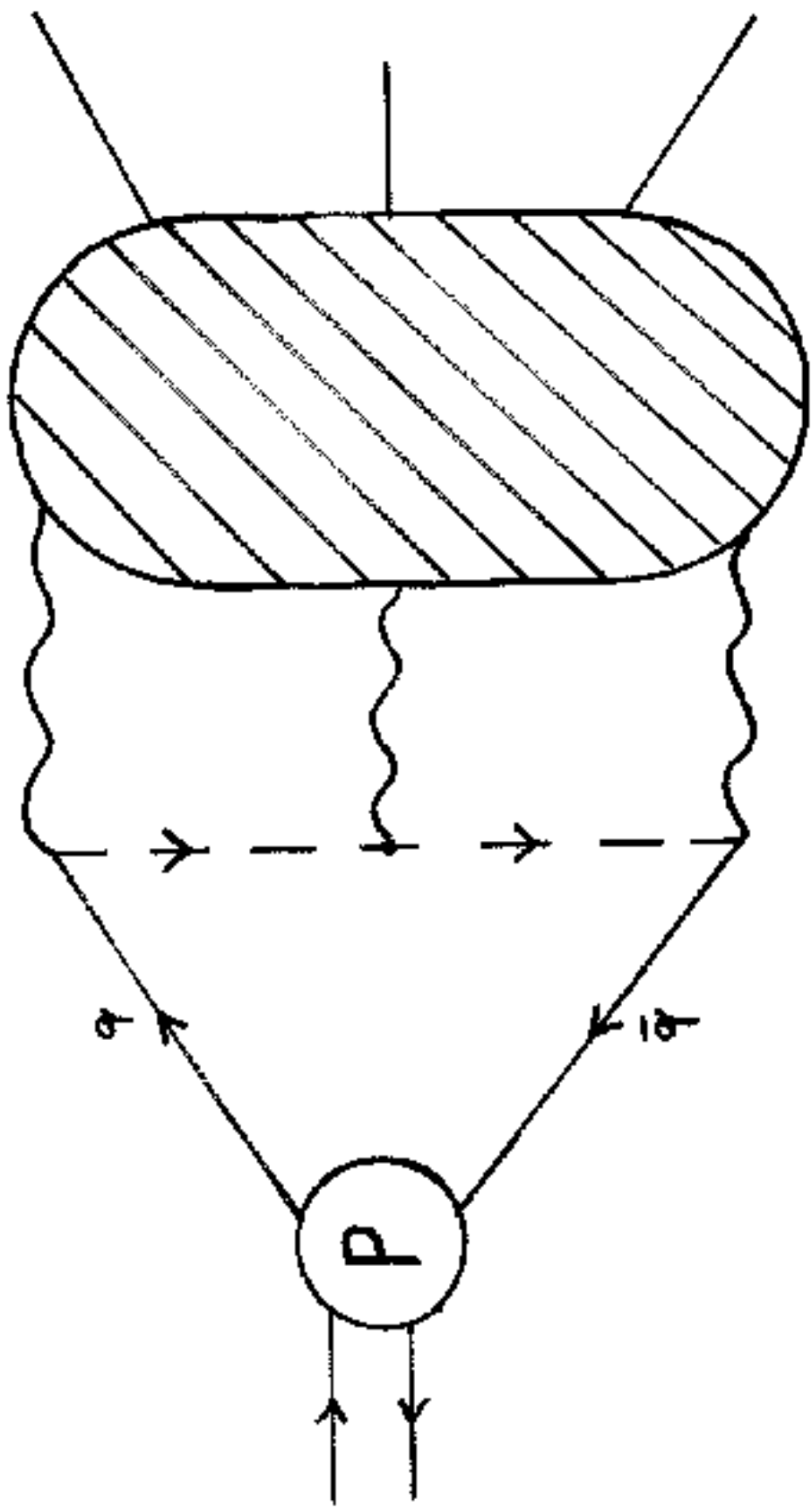
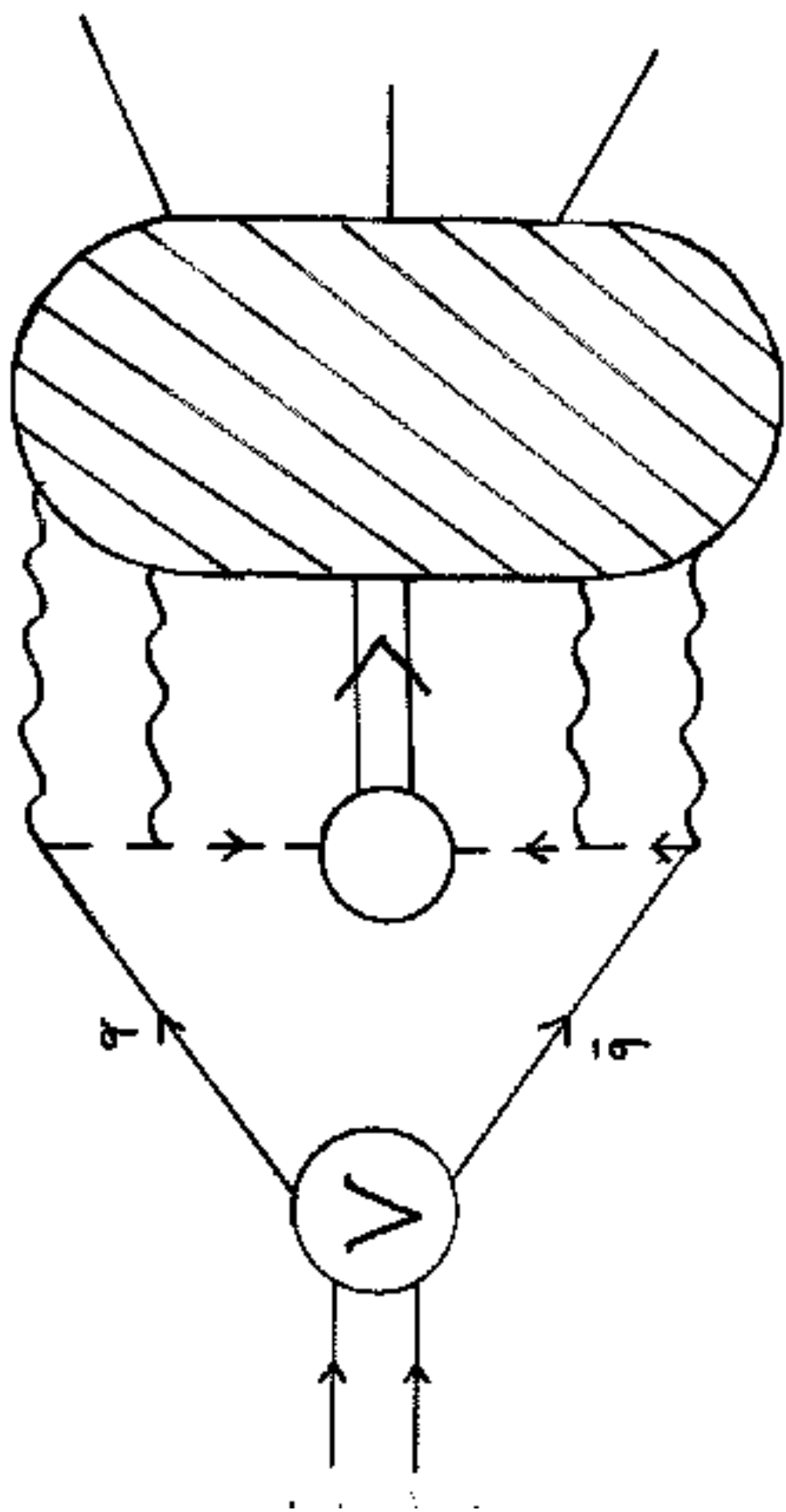
\end{document}